\documentclass[showpacs, twocolumn]{revtex4}
\usepackage{amsfonts}
\usepackage{amsmath}
\usepackage{amssymb}
\usepackage{graphicx}

\begin{document}

\title{Atom walking in a traveling-wave light}
\author{Wenxi Lai}\email{wxlai@pku.edu.cn}
\affiliation{School of Applied Science, Beijing Information Science and Technology University, Beijing 100192, China}

\begin{abstract}
In this paper, we investigate mechanical motion of ultra-slow single atoms considering each atom is coherently coupled to a traveling-wave light. The main noise in this system is originated from Doppler broadening due to the continuous momentum distribution in atom wave packet. Here, it is proved that the Doppler broadening could be effectively suppressed in strong coupling regime. Under the coherent coupling, individual neutral atoms periodically walk in a definite direction. Direction of the motion depends on occupation of the atom in its two internal states related to the optical transition, since the atom would be affected by attractive or repulsive forces depending on the internal states. It is analogous to the electric force acting on negatively or positively charged particles. We explain them with spin-orbit coupling of atoms which is hidden in our Hamiltonian. These results have potential applications for the construction of future atomic devices.
\end{abstract}

\pacs{37.10.Vz, 32.90.+a, 3.75.-b, 42.50.Ct}

\maketitle
Control and manipulation of atomic gas have wide applications in many areas. Recently, it attracts great attentions of researchers, such as atomtronic transistors~\cite{Pepino,Fuechsle,Daley,Anderson}, atomic circuits~\cite{Lee,Chow,Compagno}, atomtronic SQUID~\cite{Aghamalyan,Ryu,Jezek}, quantum flux qubits~\cite{Aghamalyan2,Safaei}, atomic batteries~\cite{Zozulya,Caliga2,Lai}, interferometries~\cite{Parazzoli,Palmer,Lachmann}, atomic clocks~\cite{Sortais,Takamoto}, precision measurements~\cite{Ido,Chapurin} and topological materials~\cite{Liu,Potirniche}.

These applications benefit from cold atom technologies which remove heat of atomic gas optically and electromagnetically~\cite{Davis,Wieman,Maunz}. Atoms in gases move more freely than they in solids. Therefore, to study behavior of individual atoms in a cold atom system is very important. It is also fundamental for the development of single atom devices~\cite{Walthe,Huesmann,Bianchet,Hetet}. Our present interest is mainly concentrated on mechanical motion of individual atoms under external fields which is at the hard of atomic transport and related applications mentioned above.

In free-space with infinite volume, a single atom could be described by a plane wave with definite momentum. However, in practical materials or devices, each atom should be described by a wave packet localized in a finite volume~\cite{Lachmann,Gattobigio}. The finite volume naturally leads to momentum uncertainty according to the Heisenberg uncertainty principle, i.e., $\Delta x \Delta p\sim h$. In atom-light interactions, the momentum uncertainty of wave packets would cause Doppler broadening to atom transition levels, which seriously influences the efficiency of control and detection of atoms~\cite{Uys,Chang}. Previously, some researchers have proposed intensity correlation method to remove Doppler broadening in atoms and molecules~\cite{Merlin}.

\begin{figure}
  \includegraphics[width=7.5 cm]{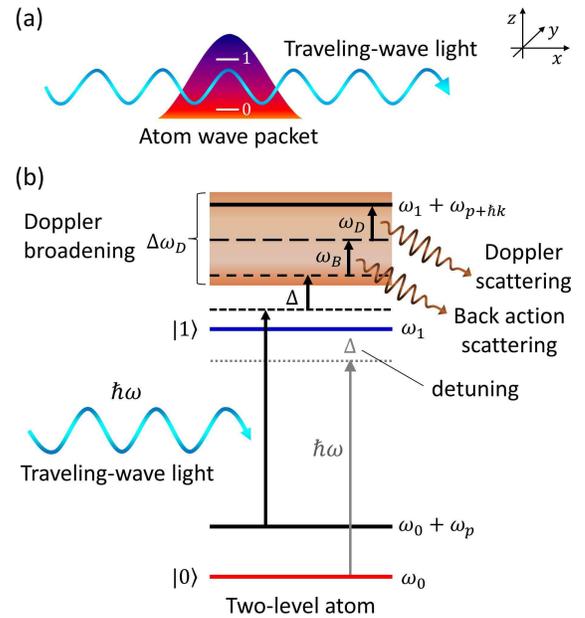}\\
  \caption{(Color on line) (a) Schematic illustration of a single atom wave packet that coupled to a traveling wave of light propagating along the $x$ coordinate. (b) Energy structure of the atom-light coupling in which Doppler effect induced energy broadening due to the continuous momentum distribution in wave packet has been considered.}
  \label{fig1}
\end{figure}

Motivated by the analogous properties between atomtronics and electronics~\cite{Pepino,Fuechsle,Daley,Anderson,Lee,Chow,Compagno,Aghamalyan,Ryu,Jezek}, in this paper, we study the motion of single atoms under a monochromatic traveling-wave light and try to exploit similarities between charged particles in electric field and neutral atoms in optical fields. To this end, here we consider ultra-slow atoms and a clean optical field. Much earlier researches on the mechanical effect of light mostly involve spontaneous emission or standing wave lasers to achieve large acceleration of atoms~\cite{Scully,Foot}. Here, taking advantage of long coherent lifetime in clock transitions of alkaline-earth atoms~\cite{Wall}, we consider pure coherent coupling between an atom wave packet and a traveling-wave light, which allow us to neglect limited lifetime of atoms. Although, Doppler shift induced noises broaden energy levels of electronic states in atom wave packet, the Doppler broadening induced noise could be suppressed in strong coupling regime as described later.

We will show that the coherent couplings make periodic oscillation of atom internal states transmit to mechanical oscillation of the atom center of mass under the drive of optical field. As a result, an individual atom could periodically walk in the light. Similar to charged particles in electric field, two-level atoms in the ground state and in the excited state move in opposite directions respectively due to the action of a light field. This behavior can be understood with the frame of spin-orbit coupling in cold atom systems~\cite{Wall,Livi,Kolkowitz}. Indeed, in the strong coupling and low momentum regime, Dirac cone could be predicted in energy structure of atom dressed states, which reflects the topologically protected locking between atom momentum and internal electronic states. Recently, spin-orbit couplings and their topological properties in cold atom systems are widely exploited with Raman transitions~\cite{Huang,Meng,Curiel,Zhang}. In the configurations of Raman transitions, sub-spaces generated through dark states play important role for the realization of coherent atom-light couplings~\cite{Zhang}.

The model in which a two-level atom wave packet interacting with a traveling-wave light is conceptually shown in Fig.~\ref{fig1} (a). The wave packet would be described by the wave function $|\Psi(t)\rangle=\int dp\sum_{n}\Psi_{n}(p,t)|n,p\rangle$ which satisfies the Schr\"{o}dinger equation $i\hbar \frac{\partial}{\partial t}|\Psi(t)\rangle=\textbf{H}|\Psi(t)\rangle$ with the total Hamiltonian $\textbf{H}=\textbf{H}_{e}+\textbf{H}_{K}+\textbf{H}_{I}$. In the Hamiltonian, the first term $\textbf{H}_{e}=\sum_{n}\hbar\omega_{n}|n\rangle\langle n|$ represents quantized energy of electronic motion around nuclear with eigenvalue $\hbar\omega_{n}$ and eigenstate $|n\rangle$. The second term describes kinetic energy of the atom wave packet in the momentum representation,
\begin{eqnarray}
&&\textbf{H}_{K}=\int d p \frac{p^{2}}{2M}|p\rangle\langle p|,
\label{eq:kinetic}
\end{eqnarray}
where $M$ is atom mass and $p$ represents momentum of the atom center of mass. As an example of the two-level system, we consider  Ytterbium ($^{173}$Yb) atom with mass number $172.938208$~\cite{Sansonetti}. It has a clock transition $^{1}$S$_{0}$ $\rightarrow$ $^{3}$P$_{0}$ that characterized by long lifetime ($\sim$ $20$ s) under the resonant wave length $578$ nm~\cite{Livi}. The states $^{1}$S$_{0}$ and $^{3}$P$_{0}$ are corresponding to the ground $n=0$ and excited state $n=1$, respectively. The traveling-wave light is described by the electric wave function $\vec{\textbf{E}}(\textbf{x},t)=\vec{e}_{z}E_{0}cos(\omega t-k\textbf{x})$ in which $\textbf{x}$ has been regarded as an operator. Then, the interaction term of the atom-light coupling could be given by,
\begin{eqnarray}
\textbf{H}_{I}=-\frac{\hbar}{2}\int d p (\Omega e^{-i\omega t}|1,p+\hbar k\rangle\langle 0,p|+H.c.),
\label{eq:int-Ham}
\end{eqnarray}
where the Rabi frequency is $\Omega=\mu E_{0}/\hbar$ with dipole moment of the atom $\mu=\langle 0|e\textbf{z}|1\rangle=|\mu|e^{i\phi}$~\cite{Scully}. The initial phase would be taken to be $\phi=0$ for convenience of discussion.
In derivation of Eq.\eqref{eq:int-Ham}, the relation of momentum displacement operator $e^{\pm ik\textbf{x}}|p\rangle=|p\pm\hbar k\rangle$ has been used~\cite{Meystre}.

To solve the wave function $|\Psi(t)\rangle$, the original Schr\"{o}dinger equation can be written into interaction picture $i\hbar \frac{\partial}{\partial t}|\varphi(t)\rangle=\textbf{V}|\varphi(t)\rangle$ with the wave function $|\varphi(t)\rangle=e^{i\textbf{H}_{0}t/\hbar}|\Psi(t)\rangle$ and time independent Hamiltonian $\textbf{V}=e^{i\textbf{H}_{0}t/\hbar}(\textbf{H}-\textbf{H}_{0})e^{-i\textbf{H}_{0}t/\hbar}$, where the unitary transformation is generated from the matrix $\textbf{H}_{0}/\hbar=\omega_{0}|0\rangle\langle 0|+(\omega_{0}+\omega)|1\rangle\langle 1|$. The wave function $|\varphi(t)\rangle$ can be expanded with probability distribution functions as $|\varphi(t)\rangle=\int d p \sum_{n}\varphi_{n}(p,t)|n,p\rangle$. For a definite momentum $p$, the Schr\"{o}dinger equation in interaction picture would be reduced into a differential equation of $2\times2$ matrix in the sub space of dressed states $\{|0,p\rangle, |1,p+\hbar k\rangle\}$ as
\begin{eqnarray}
i\left[\begin{array}{cccc}
     \dot{\varphi}_{0,p}(t) \\
    \dot{\varphi}_{1,p+\hbar k}(t)
  \end{array}\right]=\left[\begin{array}{cccc}
     \omega_{p} & -\frac{\Omega}{2} \\
    -\frac{\Omega}{2} & \Delta +\omega_{p+\hbar k}
  \end{array}\right]\left[\begin{array}{cccc}
     \varphi_{0,p}(t) \\
    \varphi_{1,p+\hbar k}(t)
  \end{array}\right],
\label{eq:equ-motion}
\end{eqnarray}
where $\omega_{p}=\frac{p^{2}}{2M\hbar}$, $\omega_{p+\hbar k}=\frac{(p+\hbar k)^{2}}{2M\hbar}$ and the detuning $\Delta=\omega_{1}-\omega_{0}-\omega$. Eq.\eqref{eq:equ-motion} can be easily diagonalized and the eigenfrequencies are $ W_{0,1}=\frac{1}{2}(\Delta+\omega_{p+\hbar k}+\omega_{p}\mp\Sigma)$, in which the energy gap is $\Sigma=\sqrt{\delta^{2}+\Omega^{2}}$ and $\delta=\Delta+\omega_{p+\hbar k}-\omega_{p}$. Solutions of Eq.\eqref{eq:equ-motion} could be written as
\begin{eqnarray}
\varphi_{0}(p,t)=(A_{0}e^{i\frac{\Sigma t}{2}}+B_{0}e^{-i\frac{\Sigma t}{2}})e^{-\frac{i}{2}(\Delta+\omega_{p+\hbar k}+\omega_{p})t},
\label{eq:solution1}
\end{eqnarray}
\begin{eqnarray}
\varphi_{1}(p+\hbar k,t)=(A_{1}e^{i\frac{\Sigma t}{2}}+B_{1}e^{-i\frac{\Sigma t}{2}})e^{-\frac{i}{2}(\Delta+\omega_{p+\hbar k}+\omega_{p})t},
\label{eq:solution2}
\end{eqnarray}
where the coefficients are
\begin{eqnarray}
A_{0}=\frac{1}{2\Sigma}[(\Sigma+\delta)\varphi_{0}(p,0)+\Omega\varphi_{1}(p+\hbar k,0)], \notag
\end{eqnarray}
\begin{eqnarray}
B_{0}=\frac{1}{2\Sigma}[(\Sigma-\delta)\varphi_{0}(p,0)-\Omega\varphi_{1}(p+\hbar k,0)],\notag
\end{eqnarray}
\begin{eqnarray}
A_{1}=\frac{1}{2\Sigma}[(\Sigma-\delta)\varphi_{1}(p+\hbar k,0)+\Omega\varphi_{0}(p,0)] \notag
\end{eqnarray}
and
\begin{eqnarray}
B_{1}=\frac{1}{2\Sigma}[(\Sigma+\delta)\varphi_{1}(p+\hbar k,0)-\Omega\varphi_{0}(p,0)].\notag
\end{eqnarray}

Before the atom interacts with light, at $t=0$, we suppose the initial wave packet could be expanded with the eigenstates $|n,p\rangle$ of free evolution system, namely $|\varphi(0)\rangle=\int dp \sum_{n}\varphi_{n}(p,0)|n,p\rangle$. In the initial state, each probability amplitude is supposed to be a Gaussian function $\varphi_{n}(p,0)=\frac{C_{n}}{\pi^{1/4}\sqrt{\Pi}}e^{-(p-p_{c})^{2}/(2\Pi^{2})}$ with the coefficient $C_{n}$ for atom electronic state $|n\rangle$. $p_{c}$ represents center of the wave packet and $\Pi$ indicates width of the wave packet in momentum space.

\begin{figure}
  \includegraphics[width=8.5 cm]{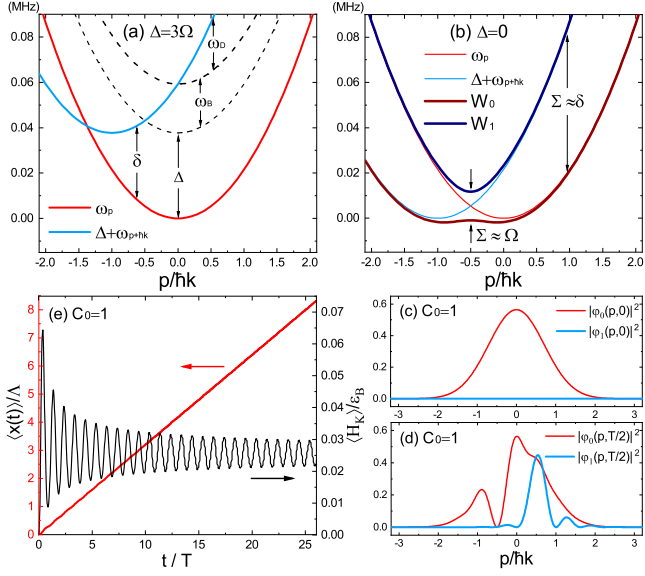}\\
  \caption{(Color on line) (a) $\omega_{p}$ and $\Delta+\omega_{p+k}$ are energy levels in the interaction picture with distance $\delta$. The dashed lines denote energy structure of $\Delta$, $\omega_{B}$ and $\omega_{D}$. (b) Comparison between eigenfrequencies and the original transition levels. $\Sigma$ is distance between two eigenfrequencies. (c) Wave packet distribution in momentum space at initial time $t=0$. (d) Wave packet distribution at time $t=T/2$. (e) Displacement $\langle x(t)\rangle$ and kinetic energy $\langle H_{k}\rangle$ of the atom as a function of time corresponding to the wave packet distribution in (e) and (d). The parameters here are $\Delta=0$, $\Omega=2\pi\times 2$ kHz, $\Pi=\hbar k$ and $p_{c}=0$.}
  \label{fig2}
\end{figure}

In the interaction picture, the modified detuning $\delta=\Delta+\omega_{p+\hbar k}-\omega_{p}$ can be rewritten into the form $\delta=\Delta+\omega_{B}+\omega_{D}$ which is sum of the detuning $\Delta$, the back action (recoil energy) shift $\omega_{B}=\frac{\hbar k^{2}}{2M}$ and the Doppler shift $\omega_{D}=\frac{Pk}{M}$ as illustrated in Fig.~\ref{fig2} (a). Corresponding back action energy is denoted as $\varepsilon_{B}=\hbar\omega_{B}$. Since the Doppler shift is proportional to momentum of the wave packet which has continuous distribution in momentum space, the optical transition levels would have the Doppler broadening $\Delta\omega_{D}=\frac{2\Pi k}{M}$ related to the width $\Pi$ of the wave packet (see Fig.~\ref{fig1} (b)). In the modified detuning $\delta$, the back action shift can be removed by taking a detuning $\Delta=-\omega_{B}$. However, the Doppler shift $\omega_{D}$ can't be removed simply by choosing a detuning $\Delta$. Because the Doppler shift can be continues values within the Doppler broadening $\Delta\omega_{D}=\frac{2\Pi k}{M}$ which is proportional to width $\Pi$ of the wave packet.

From the point of view of eigenfrequencies shown in Fig.~\ref{fig2} (b), the energy gap $\Sigma=\sqrt{\delta^{2}+\Omega^{2}}$ has two regimes, the weak coupling regime $\delta\gg\Omega$ and the strong coupling regime $\delta\ll\Omega$. In the weak coupling regime, for example, we take $\Omega=2\pi\times 1$ kHz in Fig.~\ref{fig2} (c), (d) and (e). Initially, internal state of the atom is in the ground state $C_{0}=1$ and corresponding wave packet is supposed to be a Gaussian function as plotted in Fig.~\ref{fig2} (c). After the atom-light interaction begins, as illustrated in Fig.~\ref{fig2} (d), momentum distribution of the wave packet would change along with the atom transition from the ground state to the excited state. It induces change in the averaged momentum of the atom and so on the corresponding kinetic energy as shown in Fig.~\ref{fig2} (e). However, the coupling strength $\Omega$ is very small in the weak coupling regime. As a result, the modified Rabi frequency $\Sigma=\sqrt{\delta^{2}+\Omega^{2}}$ remarkably depends on momentum $p$ due to $\delta$ is related to the momentum. After a longer time $t$, the momentum distribution in the excited state $\varphi_{1}(p,t)$ would spread more widely. Then, the wave packet could be regarded as many parts with different possible values of momentum. All these parts of the wave packet involved in the resonant interaction would oscillate with different Rabi frequencies $\Sigma$. Therefore, the averaged momentum of the atom (reflected by the kinetic energy in Fig.~\ref{fig2} (e)) increases in a definite direction at the beginning and then tends to decay into a stationary value because of the counterbalance between those resonant transitions. Therefore, in the weak coupling regime, atom waking is hard to be observed although displacement of the atom increases.

\begin{figure}
  \includegraphics[width=8.5 cm]{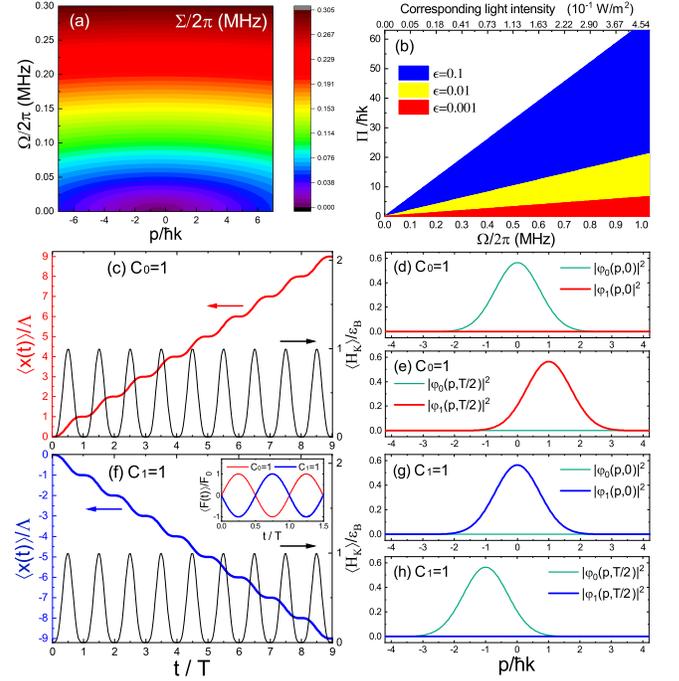}\\
  \caption{(Color on line) (a) The energy gap $\Sigma$ as a function of the Rabi frequency $\Omega$ and atom momentum $p$. (b) Illustration of strong coupling and weak coupling regimes based on Eq.\eqref{eq:criteria} with the detuning $\Delta=-\omega_{B}$. Displacement, kinetic energy of the atom as a function of time and corresponding change in the probability distribution functions with the initial state $C_{0}=1$ in (c)-(e) and with the initial state $C_{1}=1$ in (f)-(h). The inset figure in (f) shows optical forces acting on the atom for different initial states. The common parameters are $\Omega=2\pi\times 1$ MHz, $\Pi=\hbar k$, $p_{c}=0$ and $\Delta=0$.}
  \label{fig3}
\end{figure}

\begin{figure}
  \includegraphics[width=7 cm]{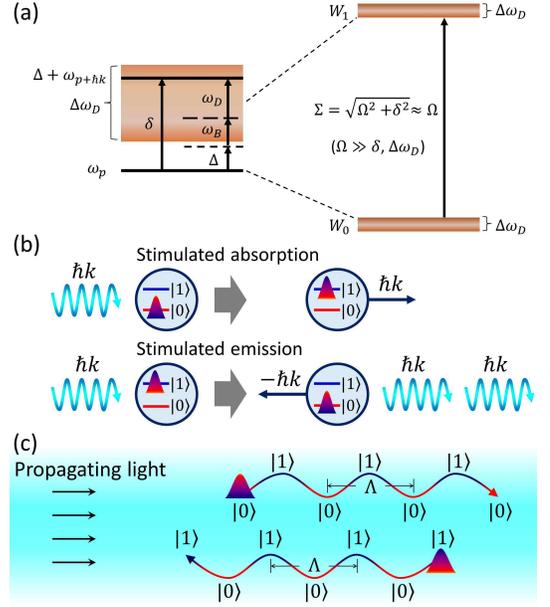}\\
  \caption{(a) The principle of strong coupling in the system. (b) Illustration of spin-orbit coupling during the atom-photon interactions. (c) Walking of atom in the opposite directions due its different initial internal states $|0\rangle$ and $|1\rangle$.}
  \label{fig4}
\end{figure}

Now, let us analyze how to identify the weak and strong coupling regimes and discuss suppression of the Doppler broadening in the latter situation. Substituting $\delta=\Delta+\omega_{B}+\omega_{D}$ into the relation $\Sigma=\sqrt{\delta^{2}+\Omega^{2}}$, one can deduce the following equation directly,
\begin{eqnarray}
\frac{(\frac{p}{\hbar k}+\frac{\Delta}{2 \omega_{B}}+\frac{1}{2})^{2}}{(\frac{\Sigma}{2 \omega_{B}})^{2}}+\frac{\Omega^{2}}{\Sigma^{2}}=1.
\label{eq:elliptic}
\end{eqnarray}
This elliptic equation could be used to analyze the two different coupling regimes. Fig.~\ref{fig3} (a) reveals the weak and strong coupling regimes are relative. For larger momentum range $-\Pi<p+\frac{\Delta}{2 \omega_{B}}+\frac{1}{2}<\Pi$, a larger Rabi frequency $\Omega$ is required to satisfy the strong coupling regime where $\delta\ll\Omega$ and $\Sigma\approx\Omega$. In this case, the Doppler shift included in $\delta$ becomes negligible small (see Fig.~\ref{fig4} (a)). This is important for the following result of atom walking. To find a quantitative relation between the width $\Pi$ of a wave packet and the Rabi frequency $\Omega$ for the strong coupling, an infinitely small parameter $\epsilon$ could be defined that $\Sigma=\Omega(1+\epsilon)$. It guarantees that $\Sigma$ is bigger than $\Omega$ but can be close to $\Omega$ infinitely. Then considering Eq.\eqref{eq:elliptic} the criteria of the Rabi frequency $\Omega$ that satisfies the strong coupling regime can be found from the equation,
\begin{eqnarray}
\Omega=\frac{\sqrt{2}\omega_{B}}{\sqrt{\epsilon}}(\frac{\Pi}{\hbar k}+\frac{\Delta}{2\omega_{B}}+\frac{1}{2}).
\label{eq:criteria}
\end{eqnarray}
In Fig.~\ref{fig3} (b), range of the Rabi frequency $\Omega$ given by Eq.\eqref{eq:criteria} for different $\epsilon$ has been plotted for the detuning $\Delta=-\omega_{B}$. Obviously, when $\epsilon$ is small enough the strong coupling should be realized.

In Fig.~\ref{fig3} (c-h), the Rabi frequency $\Omega=2\pi\times1$ MHz and the width $\Pi=\hbar k$ satisfy Eq.\eqref{eq:criteria} with parameter $\epsilon=0.001$ as revealed in Fig.~\ref{fig3}(b). In this case, almost fully periodic motions of the atom appear in Fig.~\ref{fig3} (c) and (f). Expectation value of the wave packet momentum has been calculated through the formula $\langle p(t)\rangle=\langle\varphi(t)|p|\varphi(t)\rangle$. In the strong coupling regime $\delta\ll\Omega$, it has the simple form,
\begin{eqnarray}
\langle p(t)\rangle=p_{c}+(C_{0}^{2}-C_{1}^{2})\hbar k sin^{2}(\frac{\Omega t}{2}).
\label{eq:mome}
\end{eqnarray}
Consequently, kinetic energy of the atom could be written as $\langle H_{K}\rangle=\frac{\langle p(t)\rangle^{2}}{2M}$. From the momentum $\langle p(t)\rangle$ one can obtain displacement of the atom,
\begin{eqnarray}
\langle x(t)\rangle=\langle x(0)\rangle+\frac{2p_{c}\upsilon t}{\hbar k}+(C_{0}^{2}-C_{1}^{2})\upsilon (t-\frac{sin(\frac{2\pi t}{T})}{2\pi/T}),
\label{eq:disp}
\end{eqnarray}
where the velocity $\upsilon=\frac{\Lambda}{T}$ is ratio of the step length $\Lambda=\frac{\pi\hbar k}{M\Omega}$ and the periodicity $T=\frac{2\pi}{\Omega}$. The step length $\Lambda$ of atom walking is related to momentum $\hbar k$ of a photon, atom mass $M$ and Rabi frequency. It can be tuned arbitrarily with the Rabi frequency for a given atom. Change in the momentum gives rise to optical force acting on the atom,
\begin{eqnarray}
\langle F(t)\rangle=(C_{0}^{2}-C_{1}^{2})F_{0} sin(\Omega t),
\label{eq:force}
\end{eqnarray}
where, $F_{0}=\frac{1}{2}\hbar k \Omega$ is the force amplitude.

Eq.\eqref{eq:disp} reveals that atoms in the ground state $|0\rangle$ and the excited state $|1\rangle$, which are initially static, would move in opposite directions respectively under the action of a propagating light. This phenomenon is analogous to the positively and negatively charged particles move in opposite directions respectively under the action of an electric field. It is originated from the oscillating light forces have phase difference of $\pi$ with respect to the two internal states of atom (see Fig.~\ref{fig3} (f)). Periodic motion of the atom in opposite directions shown in Fig.~\ref{fig3} (c) and (f) could be understood with the model of stimulated absorption and stimulated emission in Fig.~\ref{fig4} (b). Different from the selected transition of wave packet in Fig.~\ref{fig2} (d), wave packet in the strong coupling regime entirety transits from one state to the other state synchronously as illustrated in Fig.~\ref{fig3} (d), (e), (g) and (h). This synchronous transitions of wave packet lead to periodic walking of the atom with step length $\Lambda$ which is schematically shown in Fig.~\ref{fig4} (c). Kinetic energy of the wave packet is oscillating between $0$ and value of the back action energy as plotted in Fig.~\ref{fig3} (c) and (f). It reveals that kinetic energy of the atom comes from the back action effect.

Now, in the following three paragraphs, let us summaries relations between the atom-light coupling, the Doppler shift and atom walking. In the model, it is clear that the optical field has a very narrow frequency with negligible width. At the same, the atom wave packet has continuous spectrum of momentum and width of the spectrum can not be neglected. It means the atom moves at continuously different velocities. Therefore, when the atom encounters a monochromatic light, Doppler shifts with continuous different values occur to their energy levels. The atom would observe the light with wide frequency spectrum, or equivalently, the light couples to the atom with a remarkable level width. This is the Doppler broadening mentioned above. In this case, the atom is equivalently interacting with a noisy environment which is characterized by a continuous energy spectrum. This is why the atom damping occurs in Fig.~\ref{fig2} (e).

Without considering the Doppler shift, when a two-level atom resonantly interacts with a monochromatic light, the atom would oscillate with the definite Rabi frequency $\Omega$~\cite{Scully,Meystre}. Considering Doppler shift, but supposing the atom has only definite momentum $p$, resonant atom-light coupling would enable the atom oscillate with the new Rabi frequency $\Sigma$. The Doppler shift is included in the modified frequency $\Sigma$. When both Doppler shift and momentum spread of the atom wave packet are considered, the new Rabi frequency $\Sigma$ has continuously different values related to momentum $p$. It reveals the atom (or wave function) would oscillate with different Rabi frequency $\Sigma$. What we see in classical world is the decays happen to the oscillation of atom occupation and atom walking.

However, the decays would be removed when we takes strong coupling. There are two regimes for the atom-light interaction. The situation shown in Fig.~\ref{fig2} (c)-(e) is in the weak coupling regime, the situation shown in Fig.~\ref{fig3} (c)-(h) is in the strong coupling regime. In Fig.~\ref{fig2} (c), atom is initially ($t=0$) set in the ground state. After half periodicity ($t=T/2$), only part
of the momentum spectrum is changed. It indicates the atom oscillates at different frequencies as mentioned above. In other words, the momentum oscillation is not synchronized. In the strong coupling regime, all continuously distributed momentums oscillate at the same steps as illustrated in Fig.~\ref{fig3} (d) and (e) or (g) and (h). It is because, in the strong coupling regime, Rabi frequency is set to be much larger than any Doppler shift, $\Omega\gg\omega_{D}$, so that one reaches the approximate relation $\Sigma\approx\Omega$. In this case, the new Rabi frequency $\Sigma$ almost does not rely on the momentum distribution. As a result, the coherent oscillation of atom occupation and momentum center of mass would be observed.

\begin{figure}
  \includegraphics[width=6 cm]{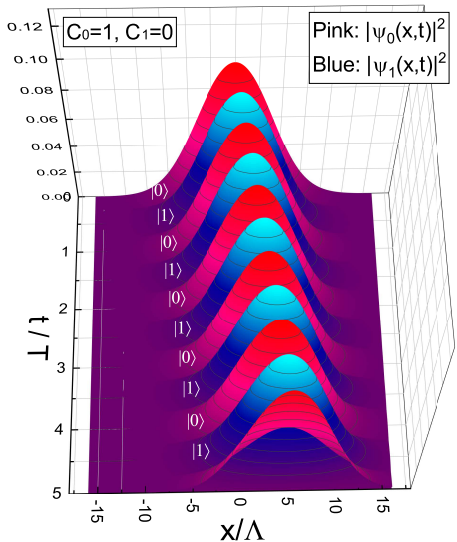}\\
  \caption{(Color on line) Atom walking along the positive direction of $x$ coordinate in real space, in which the atom is initially in its ground state. The corresponding parameters are $\Omega=2\pi\times 1$ MHz, $\Pi=50 \hbar k$, $p_{c}=0$ and $\Delta=0$.}
  \label{fig5}
\end{figure}

\begin{figure}
  \includegraphics[width=6 cm]{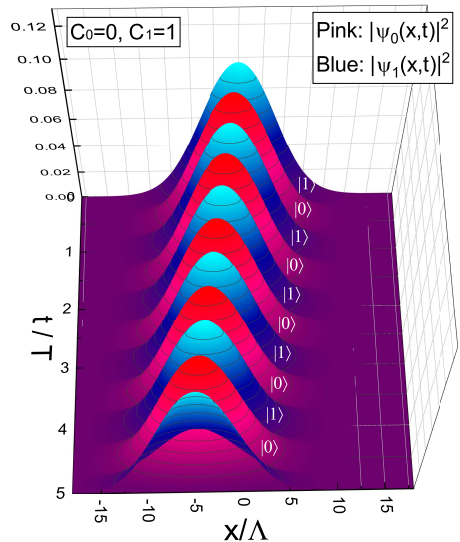}\\
  \caption{(Color on line) Atom walking along the negative direction of $x$ coordinate in real space, in which the atom is initially in its excited state. The corresponding parameters are $\Omega=2\pi\times 1$ MHz, $\Pi=50 \hbar k$, $p_{c}=0$ and $\Delta=0$.}
  \label{fig6}
\end{figure}

Next, we show atom walking in position-space. Position-space wave function of the wave packet could be written as $|\psi(x,t)\rangle=\psi_{0}(x,t)|0\rangle+\psi_{1}(x,t)|1\rangle$,
where the amplitudes are obtained through the Fourier transformation $\psi_{0}(x,t)=\int dp \varphi_{0}(p,t)\langle x|p\rangle$ and $\psi_{1}(x,t)=\int dp \varphi_{1}(p,t)\langle x|p\rangle$. The amplitudes $\varphi_{0}(p,t)$ and $\varphi_{1}(p,t)$ in momentum space are given in Eqs.\eqref{eq:solution1} and \eqref{eq:solution2}. With simple derivations one can find that Fourier transformation of the Gaussian wave function is also a Gaussian function. Fig.~\ref{fig5} show the wave packet in position space moves periodically with respect to transition of the atom between the two internal states $|0\rangle$ and $|1\rangle$. Indeed, the wave packet modes corresponding to the ground state (pink) and the excited state (blue) appear alternatively with the periodicity $T$ along the $x$ coordinate. If the atom is initially set in its excited state, it would walk in the opposite direction as illustrated in Fig.~\ref{fig6}. The wave function evolutions in Fig.~\ref{fig5} and Fig.~\ref{fig6} are equivalent to the lines of displacement $\langle x(t)\rangle$ plotted in Fig.~\ref{fig3} (c) and (f), respectively.

The walking process is similar to a longitudinal wave whose wave length is $\Lambda$ and frequency is $f=T^{-1}$. Through the concept of longitudinal wave, we reproduce the speed of atom walking given in Eq.\eqref{eq:disp}, i.e., $\upsilon=f\Lambda$. Actually, it leads to a Rabi frequency independent result $\upsilon=\frac{\hbar k}{2M}$. The speed $\upsilon$ represents average velocity of the atom walking which is exactly half of the recoil velocity obtained in the theory of atom cooling~\cite{Foot}. For the transition $^{1}$S$_{0}$ $\rightarrow$ $^{3}$P$_{0}$ of the $^{173}$Yb atom, the step length can be estimated to be $\Lambda=198.47 $ nm, $19.85$ nm and $1.98$ nm for different values of the Rabi frequency $\Omega=2\pi\times 0.01$ MHz, $2\pi\times 0.1$ MHz and $2\pi\times 1$ MHz, respectively. In this case, speed of the wave packet should be $\upsilon=1.98\times10^{6}$ nm/s.

The formula of velocity $\frac{\hbar k}{2M}$ reveals that atoms of different chemical elements or isotopes have different velocities along the direction of light propagation. The velocity which depends on mass and transition frequency of an atom could be used to separate and purify materials in principle. Details of the design for material purifications can be found in Ref.~\cite{Lai2}. It is should be noted that we just considered single atom process in the present configuration. For many-atom systems, our model is suitable to describe low density atomic gas where atom-atom interaction becomes negligible small~\cite{Gattobigio2,Couvert}.

Near the end, to further understand feature of the atom walking, let us discuss spin-orbit coupling effect and related properties included in Eq.\eqref{eq:equ-motion}. We move the origin of momentum coordinate to be $p=p_{x}-\frac{\hbar k}{2}$ in momentum space. In this case, one will find that Eq.\eqref{eq:equ-motion} becomes,
\begin{eqnarray}
i\left[\begin{array}{cccc}
     \dot{\varphi}_{0,p_{x}-\frac{\hbar k}{2}} \\
    \dot{\varphi}_{1,p_{x}+\frac{\hbar k}{2}}
  \end{array}\right]=\left[\begin{array}{cccc}
     \omega_{p_{x}-\frac{\hbar k}{2}} & -\frac{\Omega}{2} \\
    -\frac{\Omega}{2} & \omega_{p_{x}+\frac{\hbar k}{2}}
  \end{array}\right]\left[\begin{array}{cccc}
     \varphi_{0,p_{x}-\frac{\hbar k}{2}} \\
    \varphi_{1,p_{x}+\frac{\hbar k}{2}}
  \end{array}\right],
\label{eq:equ-mot-sym}
\end{eqnarray}
where, we have set the detuning $\Delta=0$ for symmetry of this equation. The Hamiltonian of the Schr\"{o}dinger equation \eqref{eq:equ-mot-sym} could be written in the form,
\begin{eqnarray}
V(p_{x})=-\upsilon p_{x}\sigma_{z}+\alpha\sigma_{x}+\beta I,
\label{eq:Ham-V}
\end{eqnarray}
where $\upsilon=\frac{\hbar k}{2M}$ denotes the mean velocity of an atom exactly given in Eq.\eqref{eq:disp}. The other two coefficients are $\alpha=-\frac{\hbar\Omega}{2}$ and $\beta=\frac{p_{x}^{2}}{2M}+\frac{\hbar^{2} k^{2}}{8M}$, respectively. Additionally, the identity matrix and the Pauli matrices represent pseudospin of the atom and they are defined to be
\begin{eqnarray}
I=\left[\begin{array}{cccc}
     1 & 0 \\
    0 & 1
  \end{array}\right], \sigma_{x}=\left[\begin{array}{cccc}
     0 & 1 \\
    1 & 0
  \end{array}\right], \sigma_{z}=\left[\begin{array}{cccc}
     1 & 0 \\
    0 & -1
  \end{array}\right].\notag
\end{eqnarray}
The first term $-\upsilon p_{x}\sigma_{z}$ on the right side of Eq. \eqref{eq:Ham-V} represents spin-orbit coupling of our system. When momentum is close to zero, $p_{x}\rightarrow 0$, the higher term $p_{x}^{2}$ decreases faster than the lower term $p_{x}$ in Eq. \eqref{eq:Ham-V}. Then, this low momentum area enables us to write the Hamiltonian as
\begin{eqnarray}
V(p_{x})\approx -\upsilon p_{x}\sigma_{z}+\alpha \sigma_{x}+\frac{\hbar^{2} k^{2}}{8M}I,
\label{eq:Ham-V-appx}
\end{eqnarray}
which is linearly proportional to the momentum $p_{x}$ and called Dirac like Hamiltonian. The spin-orbit coupling and the Dirac cone could be related to topological edge states as exploited in many recent researches on atom gases~\cite{Huang,Meng,Curiel,Zhang,Cooper}. The Dirac cone in our system appears in Fig.~\ref{fig2} (b) near the point $p=-\hbar k/2$ ($p_{x}=0$). The area near this position ($p_{x}=0$) is namely the strong coupling regime. In this regime, atoms move in different directions just depending on their internal states which is similar to the particle behavior in topological edge states~\cite{Parappurath}. The second term $\alpha\sigma_{x}$ in Eq. \eqref{eq:Ham-V-appx} reflects the energy corresponding to Zeeman field which induces Zeeman splitting and opens degeneracy at the Dirac point~\cite{Meng,Curiel}. Splitting of the Dirac point is shown in Fig.~\ref{fig2} (b). The last term is back action energy as mentioned above.

In conclusions, internal state dependent periodic motion of a single atom wave packet under the action of a traveling-wave light is demonstrated theoretically with analytic solutions of Schr\"{o}dinger equation. There are two key conditions for observation of the atom walking. One of the conditions is that atoms should have long lifetime optical transition levels, the lifetime should be at least longer than one periodicity of the atom walking. The other condition comes from the strong coupling of atom-light interaction, which suppresses the noise originated from Doppler broadening. The atom walking mainly has following properties. First, step length and time periodicity of the walking can be tuned arbitrarily by tuning the Rabi frequency. Second, direction of the atom walking is determined by initial state of the atom internal state. Therefore, the atom walking here is deterministic, it is different from the quantum random walk. Third, the atom walking has the properties of spin-orbit coupling. Forth, the highest kinetic energy of the atom during the walking process is identical to the back action energy. We believe the results in this work are important for the development of atomtronic technology.

\begin{acknowledgments}
This work was supported by R \& D  Program of Beijing Municipal Education Commission (KM202011232017).
\end{acknowledgments}

\end{document}